\documentstyle[prl,aps,psfig]{revtex} 
\begin{document}
\twocolumn[\hsize\textwidth\columnwidth\hsize\csname@twocolumnfalse\endcsname
\draft

\title{Dynamics of electrostatically-driven granular media. Effects of Humidity}
\author{D.W. Howell, I.S. Aranson, and G.W. Crabtree}
\address{Argonne National Laboratory, 9700 South Cass Avenue, Argonne, IL 60439}
\date{\today}

\maketitle

\begin{abstract}
We performed experimental studies of  
the effect of humidity on the dynamics of electrostatically-driven 
granular materials. 
Both conducting and dielectric particles  undergo a
phase transition from an immobile state (granular solid) to a 
fluidized state (granular gas) with increasing applied field.  
Spontaneous precipitation of solid clusters from the gas
phase occurs as the external driving is decreased.  
The clustering dynamics in conducting particles is 
primarily controlled by screening of the electric field 
but is aided by cohesion due to humidity.
It is shown that humidity effects 
dominate the clustering process with dielectric particles.
\end{abstract}


\narrowtext
\vskip1pc]

%Intro

Driven granular materials exhibit complex behavior that resembles some aspects of conventional 
solids, liquids, and gases, 
yet also demonstrate 
some considerable differences\cite{jnb,kadanoff,gennes,raj}.  
A key distinction that separates granular phases from their 
conventional counterparts is the presence of
dissipation of energy through inelasticity of collisions  and friction.  
%Energy must be continuously added in order to maintain granular material in a 
%fluidized  state.  
For the gas-like phase, dissipation may be responsible for non-Maxwellian
 velocity statistics\cite{menon}, and may cause the 
fluidized state to collapse locally, forming clusters of
particles at rest \cite{goldhirsh}. 
There have been some 
studies of this spatial inhomogeneity where energy is 
added by mechanically vibrating the system\cite{kudrolli,olafsen,losert}.  
Environmental effects, such as humidity and surface 
coating are additional factors 
making the physics of granular systems especially complex. 

Recent studies have explored humidity effects in static granular 
materials consisting of large particles (about 1 mm)
\cite{halsey,mason,horn,bocquet,tegzes,fraysse}, 
but little has been done to examine the  effect on dynamic systems.
This wetting effect intervenes in the
dynamics causing agglomeration, 
charging, etc.
%, making mechanical experiments uncontrollable.

Dynamics of large ensembles of small (below 30 $\mu m$) particles is even 
more challenging. There is a trend in technology to operate with  
finer particles. 
Fascinating collective behavior appears
when small particles acquire an electric charge and respond to
competing long-range electromagnetic
and short range contact forces, e.g. due to wetting. 
Unfortunately, traditional methods to probe the dynamics of 
granular materials through mechanical excitation become inefficient for 
very small particles. 
Fine particles are  more sensitive to electrostatic 
forces which arise through 
particle friction or space charges in the particle environment.  
Their large surface to volume ratio 
also amplifies the effect of water or other surfactants. 
While dry grains interact only through repulsive contact forces, wet 
grains attract each other through adhesion of liquid bridges between grains.

Recently, we studied the dynamics of granular materials consisting of 
small conducting particles through the novel technique of 
of electrostatic excitation\cite{blair}. 
Electrostatic driving makes use of these {\it bulk} forces, and  allows 
control by long-range electric forces.
This technique enables one to deal with extremely 
fine powders.
In Ref. \cite{blair} we reported on the 
discovery of phase transitions in  
electrostatically-driven granular material consisting of 
less than a single layer of conducting particles.

In this Letter we focus on the effects of controlled humidity on 
the dynamics of electrostatically driven granular materials. 
Studies were performed on both conducting and dielectric particles. 
We show that humidity does not qualitatively change the phenomenology
 of conducting particles.
%The situation is very different for 
%dielectric particles. 
In contrast, the behavior of dielectric particles changes drastically.  
In particular, 
the dielectric medium 
shows dendritic clusters at high humidity,  whereas the conducting medium
exhibits cluster coarsening irrespective of humidity. 

%Many phenomena observed in granular materials 
%involve more than one phase simultaneously. 
%This coexistence of phases and the mechanisms of 
%transitions between phases are of great interest.  

%Experimental section

\begin{figure}
\centerline{ \psfig{figure=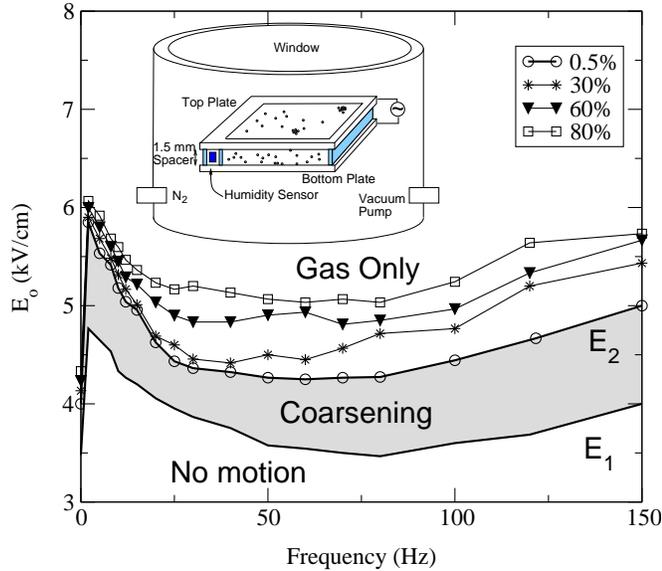,height=3in}}
\caption{
The phase diagram in 
$f,E_0$ plane for 165 $\mu$m spherical copper particles. 
Lines with symbols show $E_{2}$  for different humidity levels. 
Solid line indicates $E_1$ for 0.5\% humidity.  In {\it gas only}  
part  clusters do not form, in {\it coarsening} 
part clusters and gas coexist.  
Inset: scheme of experimental setup.
}
\label{Fig1}
\end{figure}

{\it Apparatus.} 
Our experimental setup is similar to that in Ref. \cite{blair}, although 
important modifications were made to control the moisture level inside
 the experimental cell (see Fig.\ref{Fig1}). 
Particles are placed between 
the plates of a large capacitor which is energized by 
a constant or alternating  electric field. 
To provide optical access to  the cell, the upper  
plate is made of glass with a clear conducting coating. 
We used $11\times11$ cm capacitor plates 
with a spacing of 1.5 mm.  The particles consisted of
165 $\pm$ 15 $\mu$m and 35 $\pm$ 5 $\mu$m conducting copper spheres and
165 $\pm$ 15 $\mu$m and 85 $\pm$ 5 $\mu$m glass spheres.
The field amplitude varied from 0 to 10 kV/cm
and the frequencies on the interval of 0 to 150 Hz. 
The number of particles in the cell varied between $10^5$ and $10^6$.

To control the  humidity of the system,  
the amount of water vapor was held constant by placing the
apparatus in a sealed chamber.  Before starting the experiments, 
the chamber was evacuated to approximately $10^{-6}$ torr.  
To maintain low humidity at atmospheric pressure, 
we backfilled the chamber with dry nitrogen.  
Higher humidity levels were set by feeding water vapor into the chamber, 
and allowing the system to equilibrate for $\sim 24$ hrs.  
We measured relative humidity with a capacitive sensor 
whose dielectric layer equilibrates with the surrounding gas \cite{Newark}.  
The sensor has a linear response from 0 to 90 percent humidity, 
and when calibrated with the local temperature, the accuracy of this 
measurement is $\sim 2\%$.  Since humidity between the plates 
of the experimental cell can be somewhat different 
from the surrounding chamber, the sensor must be placed within 
the apparatus to acquire an accurate measure.

%Charging of the particles results in electric current. 
%For the value of $E$ about $2E_1$
%it achieves   $1$ $\mu$a. 

{\it Conducting particles} 
acquire a surface  
charge when they are 
in contact with the 
capacitor plate. 
As the magnitude of the electric field 
in the capacitor exceeds the 
critical value $E_1$ the resulting (upward) electric 
force overcomes gravity $mg$ ($m$ is the mass of the particle, 
$g$ is the acceleration due to gravity) 
and pushes  the charged particles upward.   
When the grains hit the upper plate, they deposit  their charge and fall 
back. 
By applying an alternating electric field
$E=E_0 \sin (2 \pi f  t)$, and adjusting its frequency $f$,
one can control the vertical excursion of particles by effectively
turning them back before they collide with the upper plate. 

\begin{figure}
\centerline{ \psfig{figure=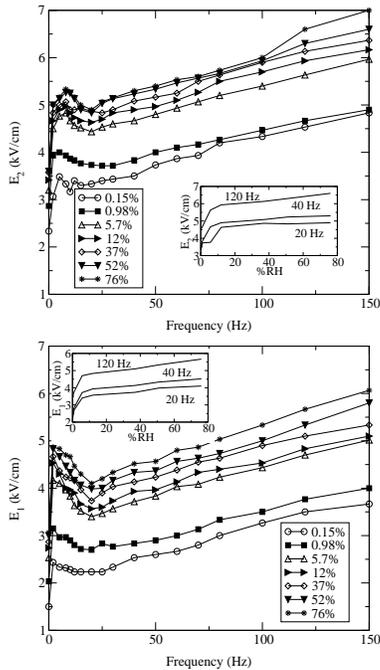,height=3.5in}}
\caption{ top: $E_{2}$ vs  
$f$ plane for several different humidities.  bottom: $E_{1}$ vs  
$f$ plane for several different humidities.
Particles are spherical copper with 35 $\mu$m diameter. 
Inset2: $E_{1,2}$ vs humidity for different values of the frequency $f$.}
\label{Fig2}
\end{figure}

The phase diagram for 165 $\mu$m copper particles is shown in Fig. \ref{Fig1}. 
%Isolated particles start to move at $E>E_1$. 
At amplitudes of the electric field 
above a {\it second} threshold value, $E_2>E_1$, the granular medium 
forms a uniform gas-like phase (granular gas). 
This second field $E_2$ is 50-70\% larger  than  $E_1$ in nearly the
whole range of the parameters used. This effect 
is also seen in the finer 35 $\mu$m copper particles 
(see Ref. \cite{blair} and Fig. \ref{Fig2}).  In the field interval $E_1<E<E_2$,
a phenomenon analogous to coalescence dynamics 
in systems exhibiting first order phase transitions \cite{coales,lp} is 
observed.  Upon decreasing the field below 
$E_2$, the gas phase loses 
its stability and small clusters of immobile  particles
surrounded by the granular gas   
form. 
These clusters then evolve via coarsening dynamics: 
small clusters disappear and large clusters grow. Eventually the 
clusters 
assume almost perfect circular form. 
A close-up image of one of the clusters is shown in Fig.\ref{Fig4}a. 
After a very large 
time ($t\approx 30000$ sec)  a single cluster containing about 
$10^5$ grains survives. 
At the final stage, a dynamic equilibrium between the
granular solid and the surrounding gas 
persists -- not all the particles join the last cluster.

We observed a very sharp peak in $E_{1,2}$ for the frequency range from 
$0<f <2$ Hz. In this range, the  typical time for cluster formation
 becomes smaller than the applied field frequency. In this frequency 
range coarsening is not well defined because the system follows the 
 instantaneous value of the electric field and samples through both 
phases each cycle. 
Similar behavior was observed  for 
small (35 $\mu$m particles, see Ref. \cite{blair}), with the difference  
that the  peak occurred 
at $f \sim 12$ Hz. This can be attributed to higher mobility of small particles
which results in a smaller time scale for cluster formation. 

The mechanism that allows cluster formation involves 
screening of the electric field by groups of particles \cite{blair}.  
If spheres are in contact with each other, 
the surface charge will redistribute: 
each sphere in the cluster 
acquires smaller charge  
than that of an 
individual sphere
due to a screening of the 
field by its neighbors. 
Details about the necessary amplitude of the applied electric 
field and the coarsening dynamics for 35 $\mu$m particles
are discussed elsewhere\cite{blair}.  
Data taken using 165 $\mu$m particles was consistent with
these previous calculations.

%Humidity effects

We are interested in how humidity affects dynamics in the system.  
Figs. \ref{Fig1},\ref{Fig2} show the $E_{1,2}$ line response for different 
humidity levels and for different particle diameters.  
The difference between particles 
in vacuum and those at atmospheric pressure with very 
low humidities was negligible.  As the amount of water vapor 
increases, the clustering starts for higher values of $E_0$ 
independent of the frequency. The same is true for the $E_1$ line. 
Our measurements indicate an almost linear increase of both critical fields $E_{1,2}$ for larger particles (165 $\mu$m) and a very abrupt increase followed by a linear regime for smaller particles (35 $\mu$m), see Insets to 
Fig. \ref{Fig2}. 

The most natural  reason for this increase
is the addition of a cohesive 
force due to humidity.  Water clings to the particles and the 
plates forming a liquid layer on the plates and around each grain.  
When these layers interact they create liquid bridges, which 
increase both the particle-plate and the particle-particle cohesion.  
This extra force prevents particles from moving at lower fields. 
Since smaller particles have larger surface/volume ratios, one expects 
the effects of surfactants to be more pronounced with small particles. 

%Material properties

%Now we turn to the effects of humidity on dielectric materials.  We 
%first discuss how dielectric particles at no humidity interact 
%with an applied electric field, and then explore effects of added water vapor.

\begin{figure}
\centerline{ \psfig{figure=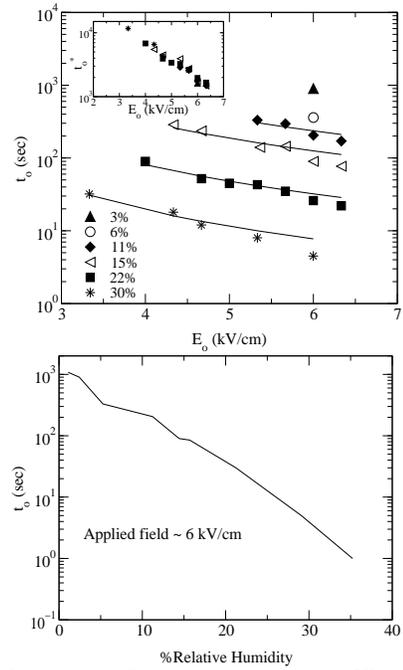,height=3.5in}}
\caption{Time before first movement $t_0$ for 85 $\mu$ glass beads. 
(a) $t_0$  vs. applied DC electric field 
Solid lines show fits according to Eq. (\protect\ref{eq2}) with
 $\tilde E=2.2$ kV/cm
and $\tau = 52, 215, 838$ and $1579$ for the humidity levels  $h=$ 
30 \%, 22 \%, 15 \% and 11 \% correspondingly. Inset:  $t_0$ vs $E_0$ 
scaled by  the factor $\exp [ \lambda h  ] $, where $\lambda=19$. 
(b) $t_0$  vs. 
Percent relative humidity for applied DC electric field of 6 kV/cm.   
}
\label{Fig3}
\end{figure}

\begin{figure}
\centerline{ \psfig{figure=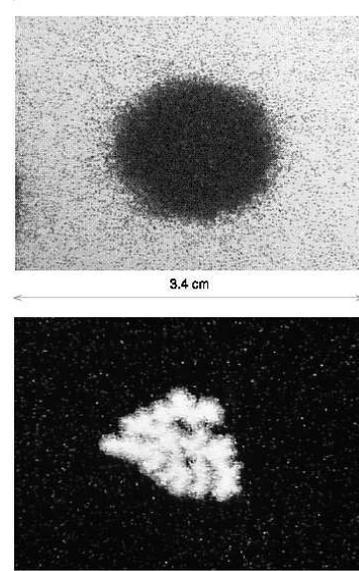,height=3in}}
\caption{
Comparison of clusters formed from copper and glass 165 $\mu$m particles.}
\label{Fig4}
\end{figure}

{\it Dielectric particles} are by definition nonconducting, 
and therefore should not be affected by uniform electric fields.  
However, materials that we consider to be electric insulators 
in practice are never perfect, and therefore, have some small 
conductivity.  The resistivity of silica glass, for example, 
is not infinite but $\sim 10^{18}$ higher than that of copper.  
The electric force $F_e$ necessary to move a dielectric particle is 
the same as that to move a conducting particle of the same weight, 
however the time it takes for a glass particle to gain enough charge for 
this force balance is determined by the time constant $\tau \sim RC$, 
where $R$ and $C$ are the effective resistance and capacitance of the 
sphere.  
Following the method used in Ref.\cite{blair}, 
the dependence of the electric force between a conducting 
sphere and a plate
is $F_e=c a^2 E^2$, where $a$ is the radius of the sphere and $E$ 
is the field in the capacitor. 
The constant $c=\zeta(3)+1/6 \approx 1.36$ comes from summing of infinite
series derived in \cite{sphere}.
If we take into account the charging process
of the sphere, the electric force assumes the form 
(compare to the charging of a capacitor): 
\begin{equation} 
F_e=  c a^2 E^2 (1 - e^{-t/\tau})
\label{eq1}
\end{equation} 
The electric force $F_e$ has to counterbalance the gravitational 
force $G=4/3 \pi \rho g a^3$, where $\rho$ is the density of the material. 
Solving for $\tau$  one derives:
\begin{equation} 
\tau = -t_0/ \log{(1- \frac{\tilde E^2}{ E^2})}
\label{eq2}
\end{equation} 
where $t_0$ is the time for first jump. Here 
$\tilde E\approx  \sqrt{ 4 \pi \rho g a /3 c }$ 
is in fact the first critical field for the conducting sphere. 
For 
85 $\mu$m  silica
glass particles ($\rho = 2.55$ $g/cm^3$) one has an estimate 
$\tilde E=2.1$ kV/cm.   
In general, we expect that the value of 
$\tilde E$ will be slightly larger due to additional cohesion forces. 
We find $t_0$ 
at
low humidity and applied field of 6 kV/cm to be 
$\sim 10^3$ sec.  
The value of $\tau$ for different values of humidity is given in Fig. 
\ref{Fig3}.

Fig. \ref{Fig3}b indicates that the dependence of $t_0$ vs the humidity 
level $h$ is in fact exponential. Indeed, all the measurements of  $t_0$
vs $E_0$ for different values of $h$ collapse on one line 
by  scaling $t_0^* = t_0/\exp(\lambda h)$ with $\lambda=19$. This exponential
(as opposed to linear) dependence of the charging time vs humidity 
is a strong indication of a probabilistic nature of the charging 
process. 
It  may resemble features of the ``asperity'' regime of Refs. \cite{halsey,mason}
of cohesion between two grains. According to Refs. \cite{halsey,mason}
the cohesion process is dominated by random 
surface roughness of the grains if the 
thickness of the fluid layer is smaller than the typical roughness scale. Since we are working at extremely small liquid fractions, we expect that 
the fluid does not cover the bead uniformly. In contrast, only a very small
fraction  of the grain  surface contributes to the charging, and this rare 
event is described by an exponential distribution.

As in the case of conducting particles, 
a surface layer of water causes glass particles to become
more cohesive, but also allows faster charging of the bead surface.  
At low humidities, 
the beads charge slowly, moving with a time near $\tau$ 
for a DC charge.  Because of the slow build up and 
release of charge, subsequent AC driving moves the particles
rapidly allowing some discharge every time the plate is touched.  
This transient AC-driven movement depends on charging time and the 
driving frequency. As the humidity is increased, 
a layer of water surrounds the particles allowing the particles 
to gain and lose charge more efficiently. 
Fig.\ref{Fig3}a shows time before the first movement $t_0$ 
vs. the applied field for different humidity levels. 
Fig.\ref{Fig3}b shows the decrease in this time for a 
specific field amplitude (6 kV/cm). 

Since the time between jump events in a dielectric at 
low humidity is of order $10^3$ sec, the formation of 
clusters due to field screening takes a prohibitively long time to observe.  
As the humidity increases, the time between events decreases, allowing
 an observable gas phase, but the subsequent clustering is dominated by a different mechanism than in the conducting case.  Added water vapor increases the probability that collisions between particles will result in cohesion. 
This leads to clusters of a different type than those 
observed with conducting grains.  Fig.\ref{Fig4} is a comparison 
between a typical screening charge-dominated metal cluster 
and a cohesion-dominated glass cluster, both consisting of 
165 $\mu$m particles.  The metal produces smooth circular shapes
(``positive surface tension''), 
while the glass produces a tower-like 
structure with many observable branches 
which is an indication of ``negative surface tension''. 
We believe that due to large charging time and strong cohesion, 
the dynamics of dielectric particles is similar to
diffusion-limited aggregation \cite{halsey3}. 
In contrast, fast charge redistribution enhances coarsening of 
clusters for conducting particles. 

{\it In conclusion}, we have shown that 
humidity has a noticeable effect on dynamics of 
electrostatically driven granular media.  In highly conducting materials,
such as copper spheres, the dynamics is manifested by coarsening which is 
due to screening of the electric field in dense cluster. 
In  this case humidity aids the process by creating cohesive 
liquid bridges between the particles and with the plate.  
In dielectric materials, both the dynamics and the 
coarsening are dominated by humidity effects.  
A water layer allows the particles to gain and lose charge more 
efficiently leading to movement, but also creates a cohesive force 
that creates clusters which behave quite differently from the conducting case.

We thank Dan Blair, Bob Behringer, Leo Kadanoff, Jerry Gollub  and Sid Nagel.
This research is supported by
US DOE,  contract \#  W-31-109-ENG-38.

\vspace{-0.5cm} 
\references

\bibitem{jnb} H.M.  Jaeger,  S.R.  Nagel,   and  R.P. Behringer, Physics
Today {\bf 49}, 32 (1996); \rmp {\bf 68}, 1259 (1996).

\bibitem{kadanoff} L.P.  Kadanoff, \rmp {\bf 71}, 435 (1999)
 \bibitem{gennes} P. G. de Gennes \rmp {\bf 71}, S374 (1999)
\bibitem{raj} J. Rajchenbach, Advances in Physics, {\bf 49}, 229 (2000)



\bibitem{goldhirsh} 
I. Goldhirsch and G. Zanetti, \prl {\bf 70}, 1619 (1993); 
S. McNamara and W. R. Young, \pre  {\bf 53}, 5089 (1996).
\bibitem{menon} F. Rouyer and N. Menon, \prl {\bf 85}, 3676 (2000)

\bibitem{kudrolli}
A. Kudrolli, M. Wolpert, and J. P. Gollub, 
\prl  {\bf 78}, 1383 (1997).   
\bibitem{olafsen} J.S. Olafsen  and J.S. Urbach,  \prl {\bf 81}, 4369 (1998).

\bibitem{losert} W.  Losert, D. G. W.  Cooper,  
and J.P. Gollub,  \pre {\bf 59}, 5855 (1999).

\bibitem{blair}  I.S. Aranson, D.  Blair, V.A.  Kalatsky, G.W.  Crabtree, 
W.-K. Kwok, V.M.  Vinokur, and U. Welp, \prl {\bf 84}, 3306 (2000);



\bibitem{halsey} T. C. Halsey and A.  J. Levine, \prl {\bf 80}, 3141 (1998) 
\bibitem{mason} T. G. Mason, A. J. Levine, D. Ertas, and T. C. Halsey, \pre 
{\bf 60}, R5044 (1999)
 
\bibitem{horn} D.J. Hornbaker, R. Albert, I.  Albert, A.-L. Barabasi,  and 
P. Schiffer, Nature {\bf 387}, 765 (1997).

\bibitem{bocquet} L. Bocquet,  E. Charlaix, S. Ciliberto, and J. Crassous, Nature {\bf 396}, 735 (1998).

\bibitem{tegzes} P. Tegzes, R. Albert, M. Paskvan, A.-L. Barabasi, T. Viscek, 
and P. Schiffer, \pre  {\bf 50}, 5823 (1999).

\bibitem{fraysse} N. Fraysse, H. Thome,  and L. Petit, Eur. Phys. J. B  {\bf 11}, 615-619 (1999).


\bibitem{Newark}  A Honeywell HIH-3096 was used to measure humidity.


\bibitem{coales} 
I.M. Lifshitz,  and V.V. Slyozov,   Zh. Eksp. Teor. Fiz. {\bf 35}, 
479 (1958) [Sov. Phys. JETP {\bf 8}, 331 (1959)].  
\bibitem{lp} 
E.M. Lifshitz,  and L.P. Pitaevsky,   {\it Physical Kinetics}
(Pergamon, London, 1981).

\bibitem{sphere}  W. Thomson, 
Phil. Mag., S. 4, {\bf 5}, N. 32, p. 267 (1853); 
Jeans, J.   {\it The Mathematical Theory of 
 Electricity  and Magnetism}, Cambridge 
University Press, Vth Ed., 1948
\bibitem{halsey3} T.C. Halsey, Physics Today {\bf 53}, 36 (2000)
\end{document}